\documentclass[journal]{IEEEtran}
\ifCLASSINFOpdf
\else
\fi
\usepackage{cases} 
\usepackage{graphicx}
\usepackage{arydshln}
\usepackage{multirow}

\makeatletter

\newcommand{\Rmnum}[1]{\expandafter\@slowromancap\romannumeral #1@}
\makeatother

\begin{document}
%
\title{Comments on ``Low-Complexity SIC Detection
Algorithms for Multiple-Input Multiple-Output
Systems"}
%
%
%

\author{Hufei~Zhu and Yanpeng~Wu
\thanks{H. Zhu is with the College of Computer Science and Software, Shenzhen University, Shenzhen 518060, China (e-mail:
zhuhufei@szu.edu.cn).}
\thanks{Y. Wu is with the Department of Information Science and Engineering, Hunan First Normal University, Changsha 410205, China (e-mail: xjzxwyp@hnfnu.edu.cn)}}

%
%

\markboth{Journal of \LaTeX\ Class Files,~Vol.~14, No.~8, August~2015}%
{Shell \MakeLowercase{\textit{et al.}}: Bare Demo of IEEEtran.cls for IEEE Journals}
%



\maketitle

\begin{abstract}
In the above paper, the
optimal-ordered successive interference cancellation (SIC)
detector proposed  for multiple
input multiple output (MIMO) systems
was claimed to require
a lower computational complexity than the
optimal-ordered SIC
detector proposed in the paper
``An Improved Square-Root Algorithm for V-BLAST Based on
Efficient Inverse Cholesky Factorization" (\emph{IEEE Trans. Wireless Commun.}, vol. 10,
no. 1, Jan. 2011), since several incorrect complexities were quoted or claimed.
In this comment, we revise the incorrect complexities, to draw the conclusion that  the above-mentioned two detectors
actually require the same dominant complexity.
\end{abstract}


%
\IEEEpeerreviewmaketitle

\section{Introduction}


\IEEEPARstart{I}{n}
 \cite{BLASTtransSP2015},
an
optimal-ordered successive interference cancellation (SIC)
detector was proposed  for multiple
input multiple output (MIMO) systems, and its
computational complexity was compared with the complexity
of the optimal-ordered SIC detector proposed in
\cite{zhfvtc2008}.
Unfortunately,  incorrect complexities  have been quoted
in \cite{BLASTtransSP2015} for the detector in \cite{zhfvtc2008} and a Givens rotation~\cite{MatrixComputationBook}, respectively,
and an incorrect complexity has been claimed for the detector in \cite{BLASTtransSP2015},
which utilizes
 a sequence of Givens rotations.  In this comment, the above-mentioned
incorrect complexities will be revised, and the corresponding conclusion on complexity comparison
will be modified.



\section{Main Remarks}

%
%
%
%
%

As in \cite{BLASTtransSP2015},  $N$ and $M$ denote the numbers of transmit and receive antennas, respectively.
Moreover,
as in  \cite{zhfvtc2008},
let
($k$, $l$) denotes
the computational complexity of $k$ complex
multiplications and $l$ complex additions.

 In
 row 5 of Table
\Rmnum{5}
in
 \cite{BLASTtransSP2015},
 only \cite{zhfvtc2008}  was cited to claim that
 the dominant worst-case complexity of  the  optimal-ordered SIC detector
 in \cite{zhfvtc2008} is
 $(\frac{1}{2}MN^2+\frac{7}{6}N^3,\frac{1}{2}MN^2+\frac{5}{6}N^3)$,
 which
 should actually be
 \begin{equation}\label{Flops2015}
\left( \frac{1}{2}MN^2+\frac{5}{6}N^3,\frac{1}{2}MN^2+\frac{1}{2}N^3\right),
 \end{equation}
 since
   (\ref{Flops2015}) has been given in
 lines 28 and 29 of the right column on \cite[p. 46]{zhfvtc2008}.
On the other hand, \cite{MatrixComputationBook}
was cited in lines 3-5 of the right column
on   \cite[p. 4630]{BLASTtransSP2015}, to claim that
a complex Givens rotation on a $(j+1)\times 2$ matrix
requires
a complexity of
\begin{equation}\label{flopsOfGivenswrong}
 (2j+2,2j+2).
 \end{equation}
  However,
  as described in
  lines 11 and 12 of the right column on
\cite[p. 46]{zhfvtc2008},
(\ref{flopsOfGivenswrong})
should be revised
into~\footnote{In equation (5.1.12) on \cite[p. 244]{MatrixComputationBook},
  the complex Givens rotation is written as $\left[ {\begin{array}{*{20}c}
   c & s  \\
   { - s^* } & c  \\
\end{array}} \right]$ with a real $c$ and a complex $s$, which is
the same  as the complex Givens rotation
 in lines 8-10 of the right column on
\cite[p. 46]{zhfvtc2008}.}
\begin{equation}\label{flopsOfGivens1928}
(3j+3,  j+1).
\end{equation}

Since the complexity of a Givens rotation has been modified from (\ref{flopsOfGivenswrong})
to  (\ref{flopsOfGivens1928}),
step $13$ in Table \Rmnum{2} of  \cite{BLASTtransSP2015} (which consists of
a sequence of  Givens rotations) should actually require
the worst-case complexity of $(\frac{1}{2}N^3, \frac{1}{6}N^3)$, instead of
$(\frac{1}{3}N^3,\frac{1}{3}N^3)$
 claimed
 in  lines 8 and 9 of the right column on
 \cite[p. 4630]{BLASTtransSP2015}.
  Accordingly, the worst-case complexity of  the  optimal-ordered SIC detector
proposed in \cite{BLASTtransSP2015}, which includes the complexity of
 the above-mentioned step $13$,
 should be  (\ref{Flops2015})
  instead of $(\frac{1}{2}MN^2+\frac{2}{3}N^3,\frac{1}{2}MN^2+\frac{2}{3}N^3)$
 claimed in
  lines 10-12 of the right column on
\cite[p. 4630]{BLASTtransSP2015}.
Thus it can be concluded that
both optimal-ordered SIC detectors proposed in \cite{BLASTtransSP2015}
and  \cite{zhfvtc2008} require the same dominant complexity,
which is ${\rm O}(MN^2+N^3)$.

\section{Numerical Experiments}
\begin{figure}[htbp]
\centering
\includegraphics[width=0.5\textwidth]{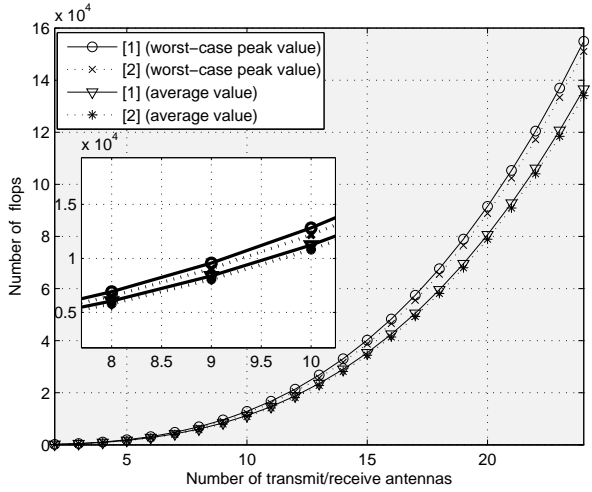}
\caption{Comparison of the worst-case and average complexities between
the two optimal-ordered SIC detectors proposed in \cite{BLASTtransSP2015}
and \cite{zhfvtc2008}.}
\end{figure}
Assume $N=M$. For different number of transmit/receive antennas, we
carried out numerical experiments
to count the worst-case and average floating-point operations (flops) of
the optimal-ordered SIC detectors proposed in
  \cite{BLASTtransSP2015} and  \cite{zhfvtc2008}, and the corresponding Matlab source code 
  with an explanatory document has been shared in \cite{ComplexityComparison1}.
  The results are shown in Fig. 1.
As in \cite{BLASTtransSP2015}
and \cite{zhfvtc2008},
the maximum number of Givens rotations are assumed to count the worst-case flops.
To count the average flops, we simulate 10000  random
channel matrices $\bf{H}$, and  neither detectors in   \cite{BLASTtransSP2015}
and \cite{zhfvtc2008} permutes the columns in $\bf{H}$ for fair comparison~\footnote{In \cite{zhfvtc2008},
the columns in the channel matrix $\bf{H}$ are permuted according to  the optimal detection order of the
adjacent subcarrier if MIMO OFDM systems are utilized, while
in \cite{BLASTtransSP2015}, the columns in $\bf{H}$ are permuted in increasing order of
their norms, or permuted equivalently by the
sorted Cholesky factorization. We do not need to compare the different methods to permute $\bf{H}$,
since the method to permute $\bf{H}$ in \cite{BLASTtransSP2015}  can be applied in
\cite{zhfvtc2008}, and vice versa. Accordingly, we do not permute  $\bf{H}$ for fair comparison.}.
From Fig. 1, it can be seen that the complexity of the detector in \cite{BLASTtransSP2015}
is close
to that of
 the detector  in \cite{zhfvtc2008}, which
is  consistent with the  complexity comparison in the last section.

\section{Conclusion}

In this comment,
we  revise
the incorrect worst-case complexity
quoted
in \cite{BLASTtransSP2015} for the optimal-ordered SIC  detector proposed in \cite{zhfvtc2008}.
On the other hand, we also
correct
the wrong complexity
quoted
in \cite{BLASTtransSP2015} for
 a Givens rotation,
 to
 revise the  worst-case
complexity claimed for the optimal-ordered SIC  detector proposed in
\cite{BLASTtransSP2015}.
By comparing the two revised worst-case complexities for the detectors in
\cite{BLASTtransSP2015} and \cite{zhfvtc2008}, we draw the conclusion that
both optimal-ordered SIC detectors proposed in \cite{BLASTtransSP2015}
and  \cite{zhfvtc2008} require the same ${\rm O}(MN^2+N^3)$ complexity,
which is then confirmed by the results of numerical experiments.

\ifCLASSOPTIONcaptionsoff
  \newpage
\fi


\begin{thebibliography}{1}

\bibitem{BLASTtransSP2015} K. Pham and K. Lee,  ``Low-Complexity SIC Detection Algorithms for
 Multiple-Input Multiple-Output Systems", \emph{IEEE
Trans. on Signal Processing}, pp. 4625-4633, vol. 63, no. 17, Sept. 2015.

\bibitem{zhfvtc2008} H. Zhu, W. Chen, B. Li, and F. Gao, ``An Improved Square-Root Algorithm for V-BLAST Based on
Efficient Inverse Cholesky Factorization", \emph{IEEE Trans. Wireless Commun.}, vol. 10,
no. 1, Jan. 2011.





\bibitem{MatrixComputationBook}
 G. H. Golub and C. F. Van Loan, \emph{Matrix Computations}, Johns Hopkins University Press,
 Baltimore, MD, 3rd edition, 1996.
 
 \bibitem{ComplexityComparison1}  Y. Wu and H. Zhu,  ``Complexity Comparison between Two Optimal-Ordered SIC MIMO Detectors Based on Matlab Simulations", arXiv preprint arXiv: 2003.03732 (2020).


%

%
%
%
%
%
%

\end{thebibliography}
\end{document}